\documentclass[preprint,showpacs,preprintnumbers,amsmath,amssymb]{revtex4}

\usepackage{graphicx}

\begin{document}

\title{
Gravitational lensing shear by an exotic lens object with
negative convergence or negative mass
}
\author{Koji Izumi}
\author{Chisaki Hagiwara}
\author{Koki Nakajima}
\author{Takao Kitamura}
\author{Hideki Asada}
\affiliation{
Faculty of Science and Technology, Hirosaki University,
Hirosaki 036-8561, Japan}

\date{\today}

\begin{abstract}
Gravitational lens models with negative convergence
(surface mass density projected onto the lens plane)
inspired by modified gravity theories, exotic matter and energy
have been recently discussed in such a way that
a static and spherically-symmetric modified spacetime metric
depends on the inverse distance to the power of positive $n$
(n=1 for Schwarzschild metric, n=2 for Ellis wormhole)
in the weak-field approximation
[Kitamura, Nakajima and Asada, PRD 87, 027501 (2013)],
and it has been shown that demagnification of images could occur
for $n>1$ lens models associated with exotic matter (and energy),
though they cause the gravitational pull on light rays.
The present paper considers gravitational lensing shear
by the demagnifying lens models and other models
such as negative-mass compact objects
causing the gravitational repulsion on light rays
like a concave lens.
It is shown that images by the lens models
for the gravitational pull are tangentially elongated,
whereas those by the repulsive ones are radially distorted.
This feature of lensed image shapes may be used for searching
(or constraining) localized exotic matter or energy
with gravitational lensing surveys.
It is suggested also that an underdense region such as a cosmic void
might produce radially elongated images of background galaxies
rather than tangential ones.
\end{abstract}


\maketitle

\section{Introduction}
The bending of light was used for
the first experimental confirmation of
the theory of general relativity.
In modern astronomy and cosmology, the gravitational lensing
is widely used, as one of the important tools,
for investigating extrasolar planets, dark matter and dark energy.

The light bending is also of theoretical importance,
in particular for studying a null structure of a spacetime.
A rigorous form of the bending angle plays an important role
in understanding properly a strong gravitational field
\cite{Frittelli, VE2000, VE2002, ERT, Perlick}.
For example,
strong gravitational lensing in a Schwarzschild black hole
was considered by Frittelli, Kling and Newman \cite{Frittelli}
and by Virbhadra and Ellis \cite{VE2000};
Virbhadra and Ellis \cite{VE2002}
and Virbhadra and Keeton \cite{VK2008}
later described
the strong gravitational lensing by naked singularities;
Eiroa, Romero and Torres \cite{ERT} treated
Reissner-Nordstr\"om black hole lensing;
Perlick \cite{Perlick} discussed the lensing
by a Barriola-Vilenkin monopole
and also that by an Ellis wormhole.

One of peculiar features of general relativity is that
the theory admits a nontrivial topology of a spacetime,
for instance a wormhole.
An Ellis wormhole is a particular example of the Morris-Thorne
traversable wormhole class \cite{Ellis, Morris1, Morris2}.
Furthermore, wormholes are inevitably related with
violations of some energy conditions in physics \cite{Visser}.
For instance, dark energy is introduced to explain
the observed accelerated expansion of the universe
by means of an additional energy-momentum component
in the right-hand side of the Einstein equation.
Furthermore, the left-hand side of the Einstein equation,
equivalently the Einstein-Hilbert action,
could be modified in various ways (nonlinear curvature terms,
higher dimensions, and so on)
inspired by string theory, loop quantum gravity and so on.
Because of the nonlinear nature of gravity,
modifications to one (or both) side of the Einstein equation
might admit spacetimes significantly different from
the standard Schwarzschild spacetime metric,
even if the spacetime is assumed to be asymptotically flat,
static and spherically symmetric.
One example is an Ellis wormhole (being an example of
traversable wormholes).

Many yeas ago, scattering problems in wormhole spacetimes were discussed
(for instance, \cite{CC, Clement}).
Interestingly, the Ellis wormhole has a zero mass
at the spatial infinity but it causes the light deflection
\cite{CC, Clement}.
Moreover, the gravitational lensing by wormholes has been recently
investigated as an observational probe of such an exotic spacetime
\cite{Safonova, Shatskii, Perlick, Nandi, Abe, Toki, Tsukamoto,
Tsukamoto2, Yoo}.
Several forms of the deflection angle by the Ellis wormhole
have been recently derived and often used
\cite{Perlick, Nandi, DS, BP, Abe, Toki, Tsukamoto}.
A reason for such differences has been clarified \cite{Nakajima, Gibbons}.

Small changes in gravitational lensing
in modified gravity theories such as $f(R)$ and fourth-order gravity
have been studied
(e.g. \cite{Capozziello, Horvath, Mendoza, Asada2011}).
Inspired by a number of works on modifications in gravitational lensing,
Kitamura et al. \cite{Kitamura} assume, in a phenomenological sense,
that an asymptotically flat, static and spherically symmetric
modified spacetime could depend on
the inverse distance to the power of positive $n$
in the weak field approximation.
The Schwarzschild spacetime and the Ellis wormhole correspond to
$n=1$ and $n=2$, respectively, so that these spacetimes
can be expressed as a one-parameter family.
Note that Birkhoff's theorem could say that cases $n \neq 1$
might be non-vacuum,
if the models were interpreted in the framework of
the standard Einstein equation.

Kitamura et al. \cite{Kitamura}
have shown that demagnification could occur for $n>1$ including
the Ellis wormhole case ($n=2$).
They have also shown that time-symmetric demagnification parts
might appear in light curves due to gravitational microlensing effects
by such exotic models.
For microlensing observations in our galaxy, light curves are available.
For cosmological situations, however, the Einstein ring size
becomes so large and hence the typical time scale is so long
that light curves cannot be observable in cosmology.
On the other hand, the image separation angle becomes sufficiently large,
so that it can be practically measured.
By using the latest result in the Sloan Digital Sky Survey Quasar Lens
Search, Takahashi and Asada have recently set the first upper bound
on the cosmic abundances of Ellis wormholes and also
negative-mass compact objects \cite{Takahashi}.
In theoretical physics, negative mass is a hypothetical concept
of matter whose mass is of opposite sign to the mass of normal matter.
Although possible negative mass ideas have been often discussed
since the 19th century, there has been no evidence for them
\cite{Bondi,Jammer1961,Jammer1999,Cramer}.
The negative masses might attract each other
to form a negative massive clump, so that
such clumps could reside in cosmological voids (e.g. \cite{Piran}).
However, the information on the image separation angle
is not sufficient for distinguishing exotic lens models.
Therefore, the main purpose of this brief paper is
to study shapes of lensed images
due to significantly modified spacetimes.

We take the units of $G=c=1$ throughout this paper.

\section{Modified spacetime model and modified lens equation}
\subsection{Modified bending angle of light}
Following Kitamura et al. \cite{Kitamura},
the present paper assumes that
an asymptotically flat, static and spherically symmetric
modified spacetime could depend on
the inverse distance to the power of positive $n$
in the weak field approximation.
We consider the light propagation through a four-dimensional spacetime,
though the whole spacetime may be higher dimensional.
The four-dimensional spacetime metric is expressed as
\begin{equation}
ds^2=-\left(1-\frac{\varepsilon_1}{r^n}\right)dt^2
+\left(1+\frac{\varepsilon_2}{r^n}\right)dr^2
+r^2(d\Theta^2+\sin^2\Theta d\phi^2)
+O(\varepsilon_1^2, \varepsilon_2^2, \varepsilon_1 \varepsilon_2) ,
\label{ds}
\end{equation}
where $r$ is the circumference radius and
$\varepsilon_1$ and $\varepsilon_2$ are small book-keeping
parameters in the following iterative calculations.
Here, $\varepsilon_1$ and $\varepsilon_2$
may be either positive or negative, respectively.
Negative $\varepsilon_1$ and $\varepsilon_2$ for $n=1$
correspond to a negative mass (in the linearized Schwarzschild metric).

Without loss of generality, we focus on
the equatorial plane $\Theta = \pi/2$,
since the spacetime is spherically symmetric.
The deflection angle of light
is obtained at the linear order as \cite{Kitamura}
\begin{align}
\alpha
&=\dfrac{\varepsilon}{b^n}\int_0^{\frac{\pi}{2}} \cos^n\psi d\psi
+O(\varepsilon^2) ,
\label{alpha}
\end{align}
where the integral is positive definite,
$b$ denotes the impact parameter of the light ray,
and we define $\varepsilon \equiv n \varepsilon_1 + \varepsilon_2$.
By absorbing the positive integral
into the parameter $\varepsilon$, we rewrite the linear-order
deflection angle simply as
$\alpha = \bar\varepsilon/b^n$,
where the sign of $\bar\varepsilon$ is the same as that of $\varepsilon$.
This deflection angle recovers
the Schwarzschild ($n=1$) and Ellis wormhole ($n=2$) cases.
For $\varepsilon > 0$, the deflection angle of light is always positive,
which means that the corresponding spacetime model causes
the gravitational pull on light rays.
For $\varepsilon < 0$, on the other hand, it is inevitably negative,
which implies the gravitational repulsion on light rays
like a concave lens.
Tsukamoto and Harada \cite{Tsukamoto2}
employ as an {\it ansatz}
the same modified bending angle as what is derived
from the spacetime metric by Kitamura et al. \cite{Kitamura}.

We mention an effective mass.
A simple application of the standard lens theory \cite{SEF}
suggests that the deflection angle of light in the form of
$\alpha = \bar\varepsilon/b^n$
corresponds to a convergence (scaled surface mass density) as
\begin{equation}
\kappa(b) = \frac{\bar\varepsilon (1-n)}{2} \frac{1}{b^{n+1}} .
\label{kappa}
\end{equation}

For the weak-field Schwarzschild case ($n = 1$),
it follows that the convergence vanishes.
For $\varepsilon > 0$ and $n>1$,
the effective surface mass density of the lens object
is interpreted as negative in the framework
of the standard lens theory \cite{Kitamura}.
This means that the matter (and energy) need to be exotic
if $\varepsilon > 0$ and $n>1$.
Also when $\varepsilon < 0$ and $n<1$,
the convergence is negative and hence
the matter (and energy) need to be exotic.
Interestingly, when $\varepsilon < 0$ and $n>1$,
the convergence is positive everywhere except for the central singularity
and hence exotic matter (and energy) are not required in the framework
of the standard lens theory, in spite of the gravitational repulsion on
light rays.
Attraction ($\varepsilon > 0$) and repulsion ($\varepsilon < 0$)
in the above models do not have a one-to-one correspondence
to positive convergence $\kappa > 0$ and negative one $\kappa < 0$.
This point is summarized in Table \ref{table-1}.

\subsection{Modified lens equation}
Under the thin lens approximation,
it is useful to consider the lens equation as \cite{SEF}
\begin{equation}
\beta = \frac{b}{D_{\rm{L}}} - \frac{D_{\rm{LS}}}{D_{\rm{S}}} \alpha(b) ,
\label{lenseq}
\end{equation}
where
$\beta$ denotes the angular position of the source and
$D_L$, $D_S$, $D_{LS}$ are the distances from the observer
to the lens, from the observer to the source, and from the lens to
the source, respectively.

For $\varepsilon > 0$,
there is always a positive root corresponding to
the Einstein ring for $\beta=0$.
The Einstein ring radius is defined as \cite{SEF}
\begin{equation}
\theta_{\rm{E}} \equiv
\left(
\frac{\bar\varepsilon D_{\rm{LS}}}{D_{\rm{S}} D_{\rm{L}}^n}
\right)^{\frac{1}{n+1}} .
\label{theta_E}
\end{equation}
If $\varepsilon < 0$, on the other hand,
Eq. (\ref{lenseq}) has no positive root for $\beta = 0$.
This is because this case describes the repulsive force.
For later convenience in normalizing the lens equation,
we define the (tentative) Einstein ring radius for $\varepsilon < 0$
as
\begin{equation}
\theta_{\rm{E}} \equiv
\left(
\frac{|\bar\varepsilon| D_{\rm{LS}}}{D_{\rm{S}} D_{\rm{L}}^n}
\right)^{\frac{1}{n+1}} ,
\label{theta_E2}
\end{equation}
though the Einstein ring does not appear for this case.
This radius gives a typical angular size for $\varepsilon < 0$ lenses.

\section{Gravitational lensing shear}
\subsection{$\varepsilon > 0$ case}
Let us begin with a $\varepsilon > 0$ case.
As already stated, the matter (and energy) need to be exotic if $n > 1$.
In the units of the Einstein ring radius,
Eq. (\ref{lenseq}) is rewritten
in the vectorial form
as
\begin{eqnarray}
\boldsymbol{\hat\beta}
&=& \boldsymbol{\hat\theta}
- \frac{\boldsymbol{\hat\theta}}{\hat\theta^{n+1}}
\quad (\hat\theta > 0) ,
\label{lenseqP}\\
\boldsymbol{\hat\beta}
&=& \boldsymbol{\hat\theta}
- \frac{\boldsymbol{\hat\theta}}{(-\hat\theta)^{n+1}}
\quad (\hat\theta < 0) ,
\label{lenseqM}
\end{eqnarray}
where we normalize $\hat\beta \equiv \beta/\theta_E$ and
$\hat\theta \equiv \theta/\theta_E$
for the angular position of the image $\theta \equiv b/D_L$,
and $\boldsymbol{\hat\beta}$ and $\boldsymbol{\hat\theta}$
denote the corresponding vectors.
There is always one image for $\hat\theta > 0$,
while the other image appears for $\hat\theta < 0$ \cite{Kitamura}.

Let us study the lensing shear
that is generally defined via the magnification matrix
$A_{ij} \equiv \partial\beta^i/\partial\theta_j$ \cite{SEF}.
After straightforward computations,
the magnification matrix for $\hat\theta > 0$
becomes explicitly
\begin{eqnarray}
(A_{ij}) &=&
\left(
\begin{array}{cc}
1 - \cfrac{1}{\hat\theta^{n+1}}
+ (n+1) \cfrac{\hat\theta_x \hat\theta_x}{\hat\theta^{n+3}}
& (n+1) \cfrac{\hat\theta_x \hat\theta_y}{\hat\theta^{n+3}} \\
(n+1) \cfrac{\hat\theta_x \hat\theta_y}{\hat\theta^{n+3}} &
1 - \cfrac{1}{\hat\theta^{n+1}}
+ (n+1) \cfrac{\hat\theta_y \hat\theta_y}{\hat\theta^{n+3}}
\end{array}
\right) .
\label{Aij}
\end{eqnarray}
It is diagonalized by using its eigen values $\lambda_{\pm}$
as
\begin{eqnarray}
(A_{ij}) &=&
\left(
\begin{array}{cc}
1-\kappa-\gamma & 0 \\
0 &  1-\kappa+\gamma
\end{array}
\right)
\nonumber\\
&\equiv&
\left(
\begin{array}{cc}
\lambda_{-} & 0 \\
0 & \lambda_{+}
\end{array}
\right) ,
\end{eqnarray}
where the $x$ and $y$ coordinates are chosen
along the radial and tangential directions, respectively,
such that $(\hat\theta_i) = (\hat\theta, 0)$
and $(\hat\beta_i) = (\hat\beta, 0)$.
Hence, the radial elongation factor is $1/\lambda_{-}$,
while the tangential one is $1/\lambda_{+}$.

First, let us investigate the primary image ($\hat\theta > 0$).
By using Eq. (\ref{lenseqP}),
we obtain
\begin{eqnarray}
\lambda_{+} = \frac{\hat\beta}{\hat\theta}
= 1 - \frac{1}{\hat\theta^{n+1}} ,
\label{lambdaP}
\end{eqnarray}
\begin{eqnarray}
\lambda_{-} = \frac{d\hat\beta}{d\hat\theta}
= 1 + \frac{n}{\hat\theta^{n+1}} .
\label{lambdaM}
\end{eqnarray}
To reach Eqs. (\ref{lambdaP}) and (\ref{lambdaM}),
we need several steps, where first the Jacobian matrix is
computed and next the matrix is diagonalized.
Note that, for our axially symmetric cases,
there is a shortcut of deriving
Eqs. (\ref{lambdaP}) and (\ref{lambdaM})
without doing such lengthy calculations.
In the shortcut, one may start with the $x$ and $y$ coordinates
that are locally chosen along the radial and tangential directions,
respectively, such that $(\hat\theta_i) = (\hat\theta, 0)$
and $(\hat\beta_i) = (\hat\beta, 0)$.
Then, infinitesimal changes in
$\boldsymbol{\hat\beta}$ and $\boldsymbol{\hat\theta}$
can be written as
$(d\hat\theta_i) = (d\hat\theta, \hat\theta d\phi)$
and
$(d\hat\beta_i) = (d\hat\beta, \hat\beta d\phi)$,
where $\phi$ denotes the azimuthal angle.
The axial symmetry allows that $\hat\theta$ and $\hat\beta$
are independent of $\phi$,
which means that the off-diagonal terms vanish
in the local coordinates.
Hence, one can immediately obtain
Eqs. (\ref{lambdaP}) and (\ref{lambdaM}) \cite{shortcut}.

\begin{figure}
\includegraphics[width=8cm]{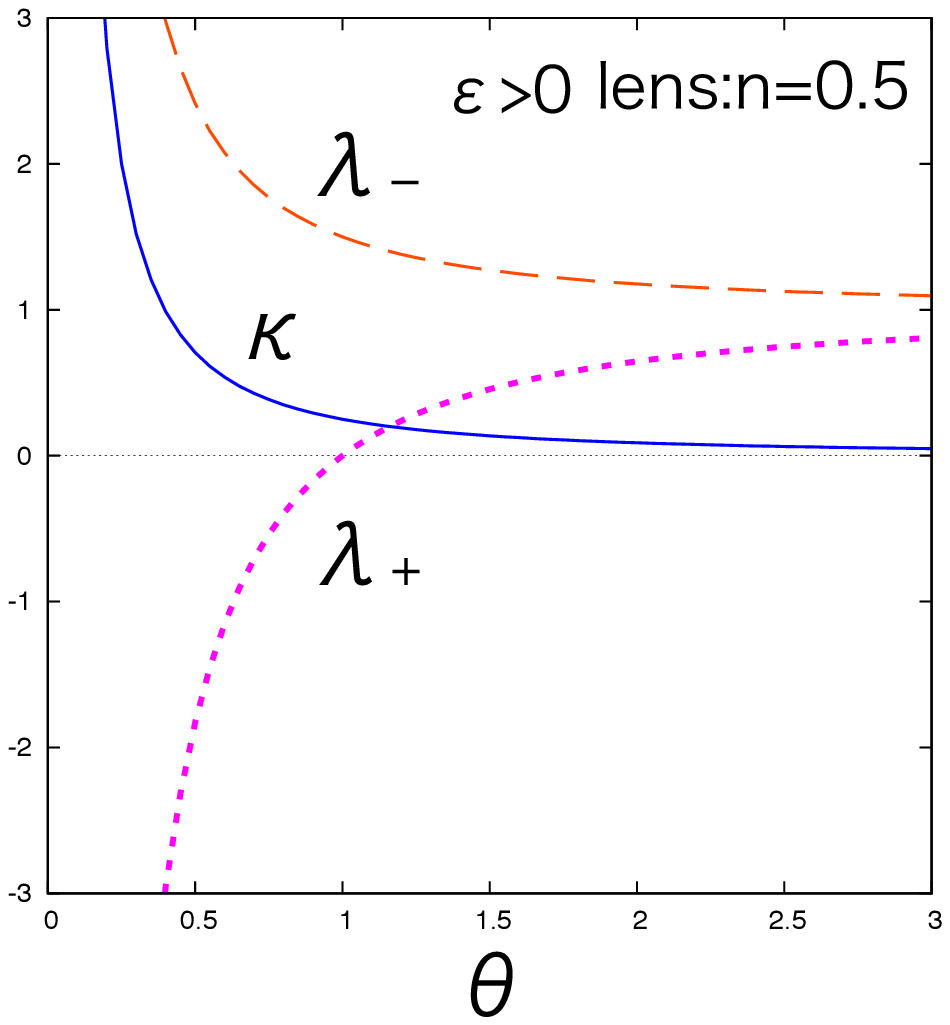}
\includegraphics[width=8cm]{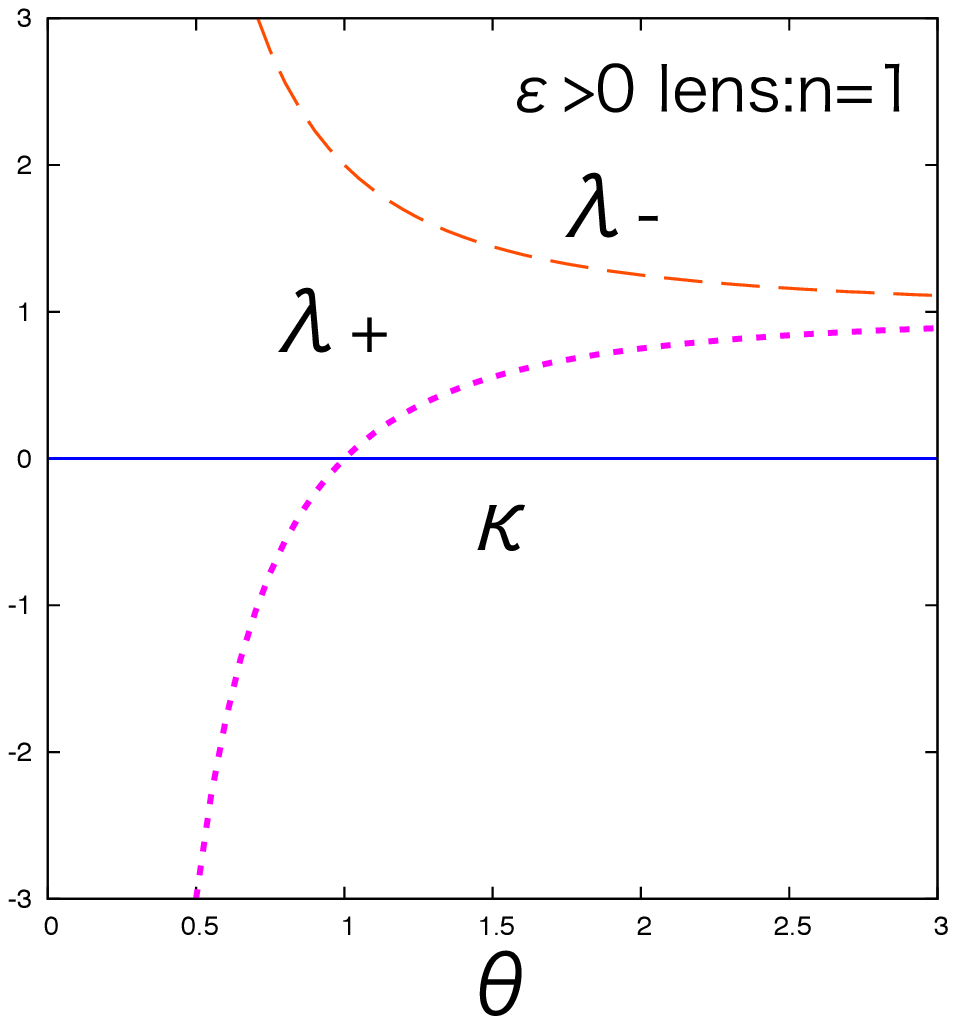}
\includegraphics[width=8cm]{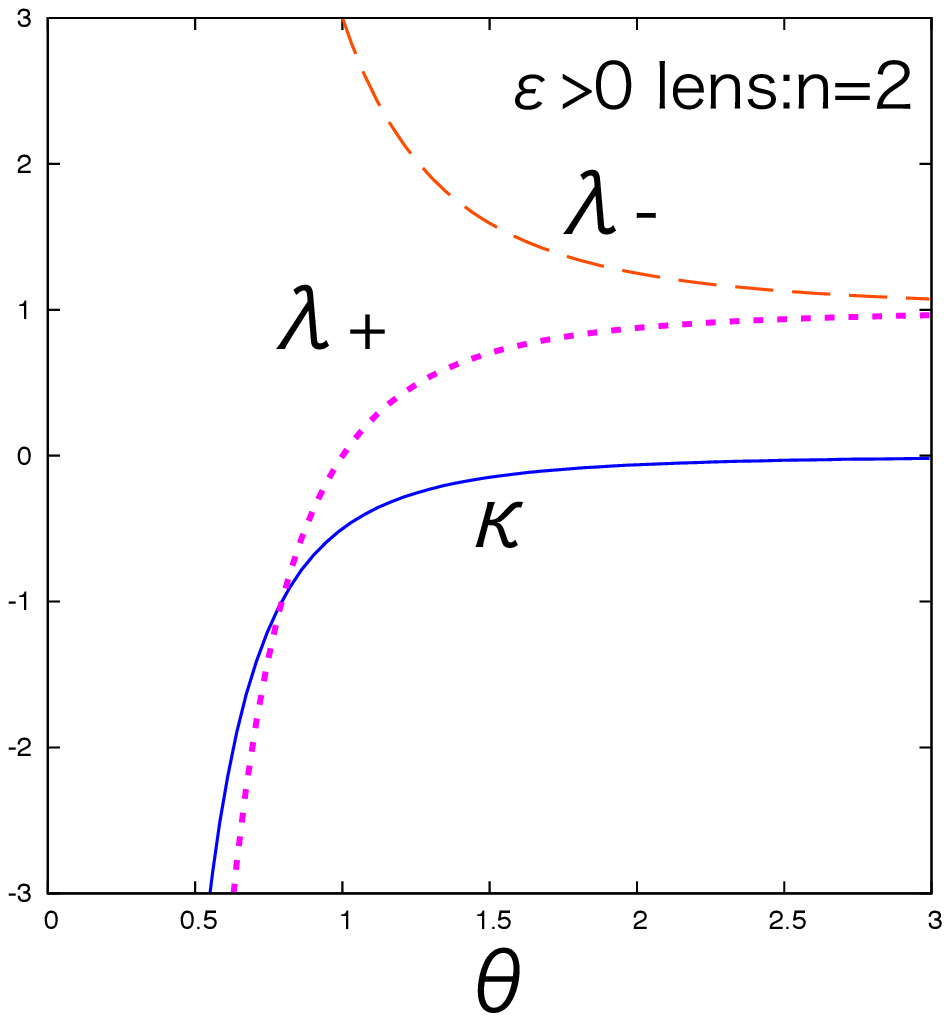}
\includegraphics[width=8cm]{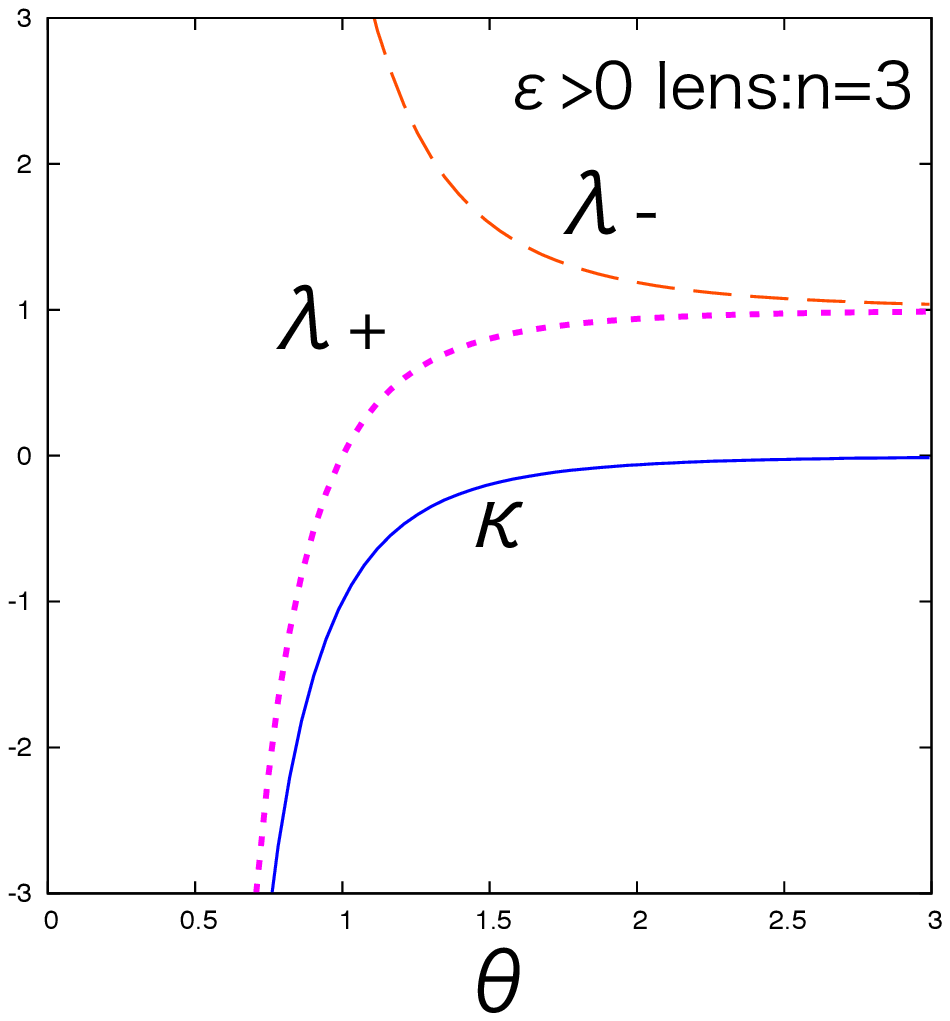}
\caption{
$\kappa$, $\lambda_{+}$ and $\lambda_{-}$ for $\varepsilon > 0$.
They are denoted by solid (blue in colors),
dotted (purple in colors) and dashed (red in colors) curves,
respectively.
The horizontal axis denotes the image position
$\theta$ in the units of the Einstein radius.
Top left: $n = 0.5$
Top right: $n = 1$.
Bottom left: $n = 2$.
Bottom right: $n = 3$.
}
\label{figure-1}
\end{figure}

If and only if $n > -1$, one can show $\lambda_{-} > \lambda_{+}$.
Therefore, the primary image is always tangentially elongated.
See also Figure \ref{figure-1} for $\kappa$ and $\lambda_{\pm}$
that are numerically calculated for $n=0.5$, $1$, $2$ and $3$.
For these four cases, $\lambda_{-}$ is always larger than $\lambda_{+}$.
The convergence $\kappa$ is positive for $n=0.5$, while
it is negative for $n=2$ and $3$.
It follows that $n=1$ corresponding to the Schwarzschild lens
leads to $\kappa = 0$.

Eqs. (\ref{lambdaP}) and (\ref{lambdaM}) give
the convergence and the shear as
\begin{eqnarray}
\kappa &=& 1 - \frac{\lambda_{+} + \lambda_{-}}{2}
\nonumber\\
&=&  \frac{1-n}{2}\frac{1}{\hat\theta^{n+1}} ,
\end{eqnarray}
\begin{eqnarray}
\gamma &=& \frac{\lambda_{+} - \lambda_{-}}{2}
\nonumber\\
&=&  - \frac{1+n}{2}\frac{1}{\hat\theta^{n+1}} ,
\end{eqnarray}
respectively.
It follows that this result of $\kappa$
agrees with Eq. (\ref{kappa}).

\begin{figure}
\includegraphics[width=8cm]{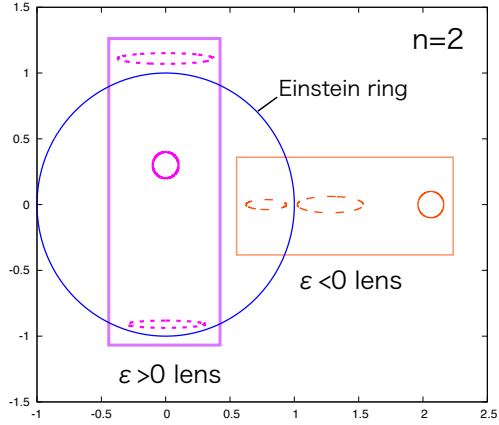}
\caption{
Numerical figures of lensed images
for attractive ($\varepsilon >0$)
and repulsive ($\varepsilon <0$) cases.
They are denoted by dashed curves.
We take $n=2$.
The source for each case is denoted by solid circles,
which are located on the horizontal axis and vertical one
for $\varepsilon < 0$ and $\varepsilon  > 0$, respectively.
}
\label{figure-2}
\end{figure}

Next, we study the secondary image ($\hat\theta < 0$).
By using Eq. (\ref{lenseqM}),
one can show $\lambda_{-} > \lambda_{+}$, if and only if $n > -1$.
Hence, the secondary image also is tangentially elongated.
See also Figure \ref{figure-2} for $\varepsilon >0$ and $n=2$,
where one can see a pair of tangential images.

Finally, we mention the dependence on the exponent $n$.
A significantly elongated case such as a giant arc
appears near the Einstein ring ($\hat\theta \sim 1$),
around which Eqs. (\ref{lambdaP}) and (\ref{lambdaM})
are expanded as
\begin{eqnarray}
\lambda_{+} &=&
(n+1) (\hat\theta - 1)
- \frac{(n+1)(n+2)}{2} (\hat\theta - 1)^2
+ O\left((\hat\theta - 1)^3\right) ,
\\
\lambda_{-} &=& n+1 -n(n+1) (\hat\theta - 1)
+ O\left((\hat\theta - 1)^2\right).
\end{eqnarray}
where we used the identity $\hat\theta = 1+(\hat\theta - 1)$.
The ratio of the tangential elongation to the radial one
(corresponding to the arc shape) is
\begin{equation}
\frac{\lambda_{-}}{\lambda_{+}}
= \frac{1}{\hat\theta - 1}
+ \left( 1 - \frac{n}{2} \right)
+ O(\hat\theta - 1) .
\end{equation}
This suggests that, for the fixed observed lens position $\hat\theta$,
elongation of images becomes weaker, when $n$ becomes larger.
This dependence on $n$ is true of also the secondary image.

\subsection{$\varepsilon < 0$ case}
\begin{figure}
\includegraphics[width=8cm]{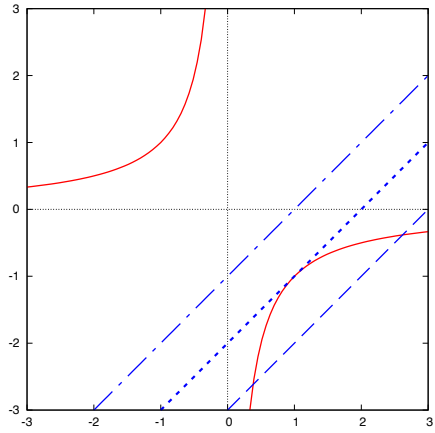}
\caption{
Repulsive lens model $(\varepsilon < 0)$.
Solid curves denote $1/\hat\theta^n$ and
straight lines mean $\hat\theta - \hat\beta$.
Their intersections correspond to image positions
that are roots for the lens equation.
There are three cases: No image for a small $\hat\beta$
(dot-dashed line),
a single image for a particular $\hat\beta$
(dotted line),
and two images for a large $\hat\beta$
(dashed line).
The two images are on the same side of the lens object.
}
\label{figure-3}
\end{figure}

Let us study $\varepsilon < 0$ case.
In the units of the Einstein ring radius,
Eq. (\ref{lenseq}) is rewritten
in the vectorial form
as
\begin{eqnarray}
\boldsymbol{\hat\beta}
&=& \boldsymbol{\hat\theta}
+ \frac{\boldsymbol{\hat\theta}}{\hat\theta^{n+1}}
\quad (\hat\theta > 0) ,
\label{lenseqP2}\\
\boldsymbol{\hat\beta}
&=& \boldsymbol{\hat\theta}
+ \frac{\boldsymbol{\hat\theta}}{(-\hat\theta)^{n+1}}
\quad (\hat\theta < 0) .
\label{lenseqM2}
\end{eqnarray}
Without loss of generality, we assume $\hat\beta >0$.
Then, Eq. (\ref{lenseqM2}) has no root satisfying $\hat\theta < 0$, while
Eq. (\ref{lenseqP2}) has at most two positive roots.
Figure \ref{figure-3} shows that there are three cases of the image
number.
For a large impact parameter case, two images appear on the same side
with respect to the lens position,
while no image appears for a small impact parameter.
The only one image appears only when the impact parameter takes
a particular value.
Let us focus on the two image cases, from which the single image case
can be discussed in the limit as the impact parameter
approaches the particular value.

\begin{figure}
\includegraphics[width=8cm]{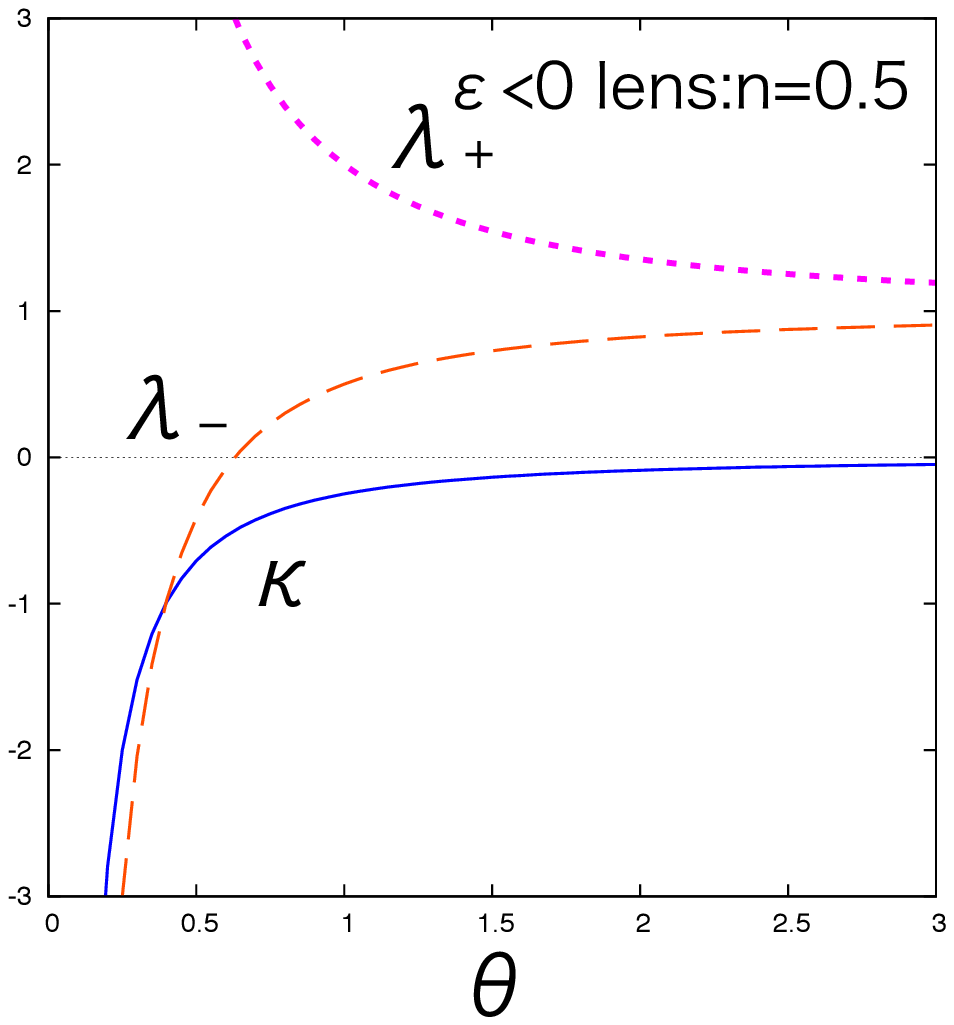}
\includegraphics[width=8cm]{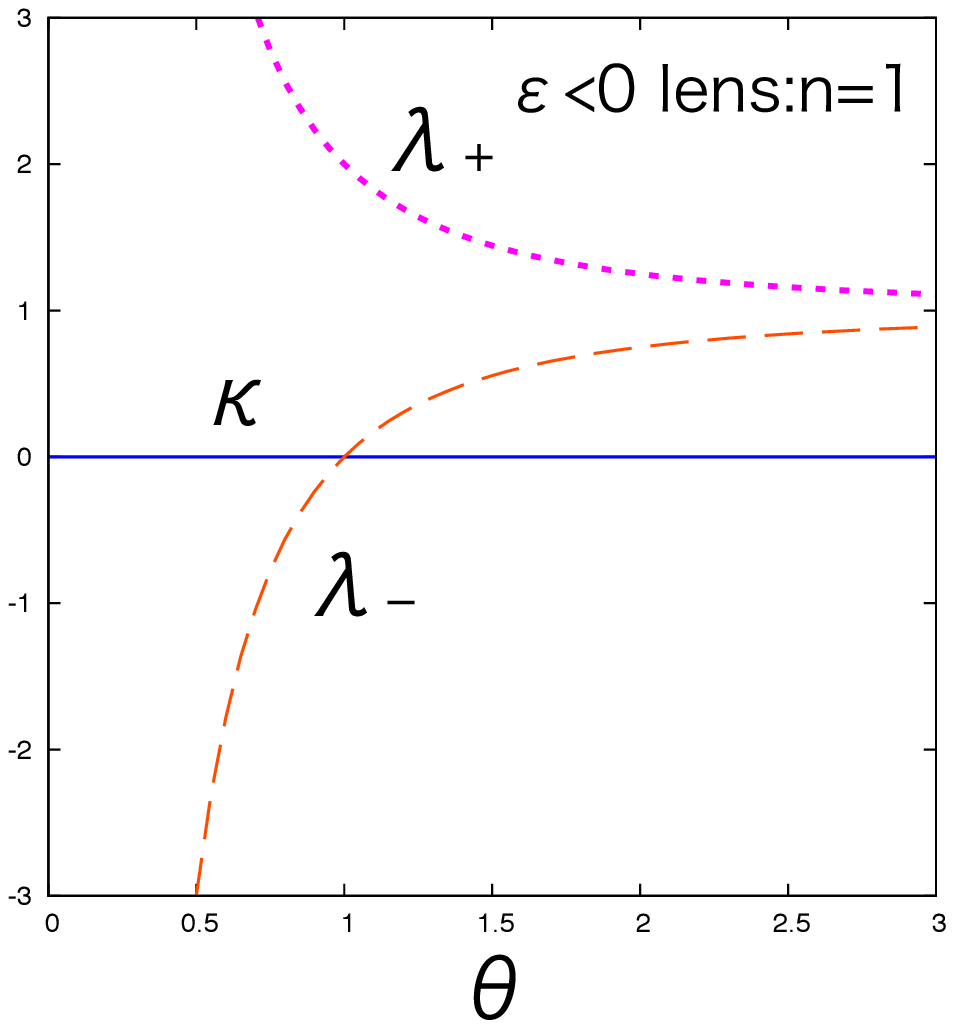}
\includegraphics[width=8cm]{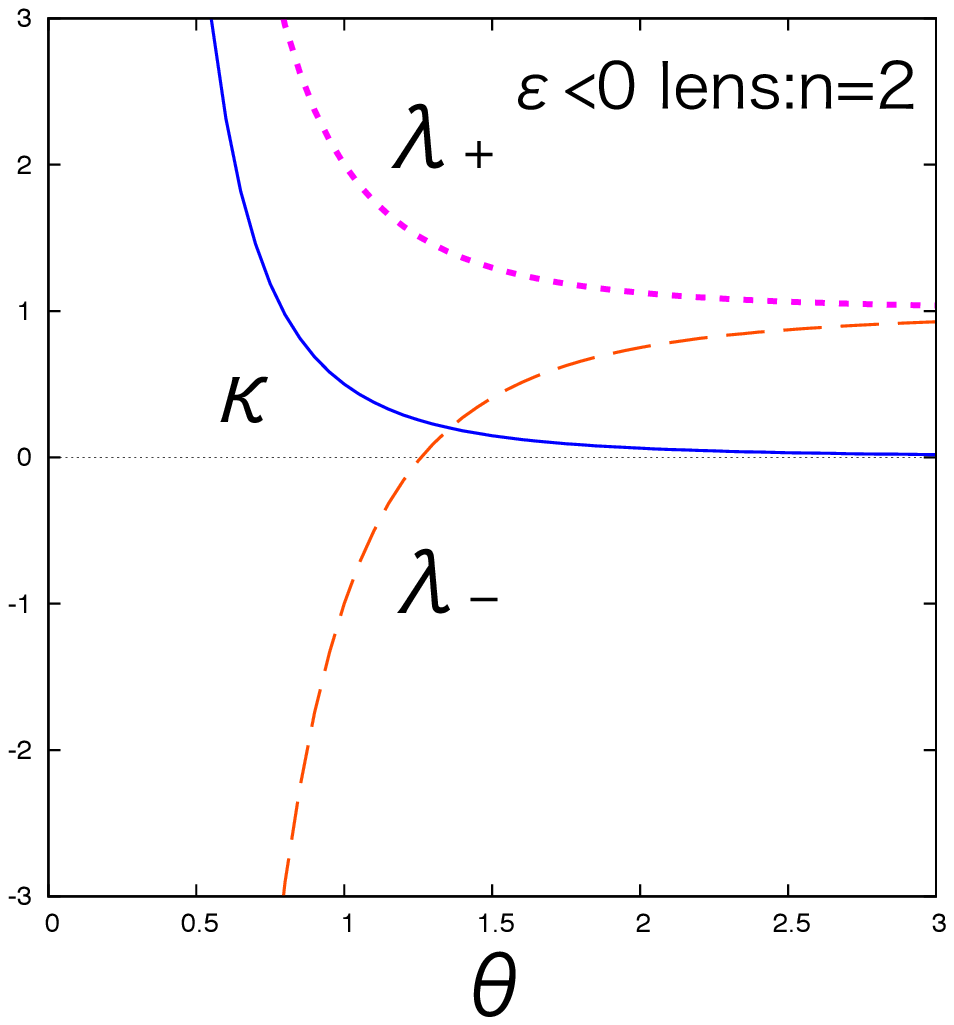}
\includegraphics[width=8cm]{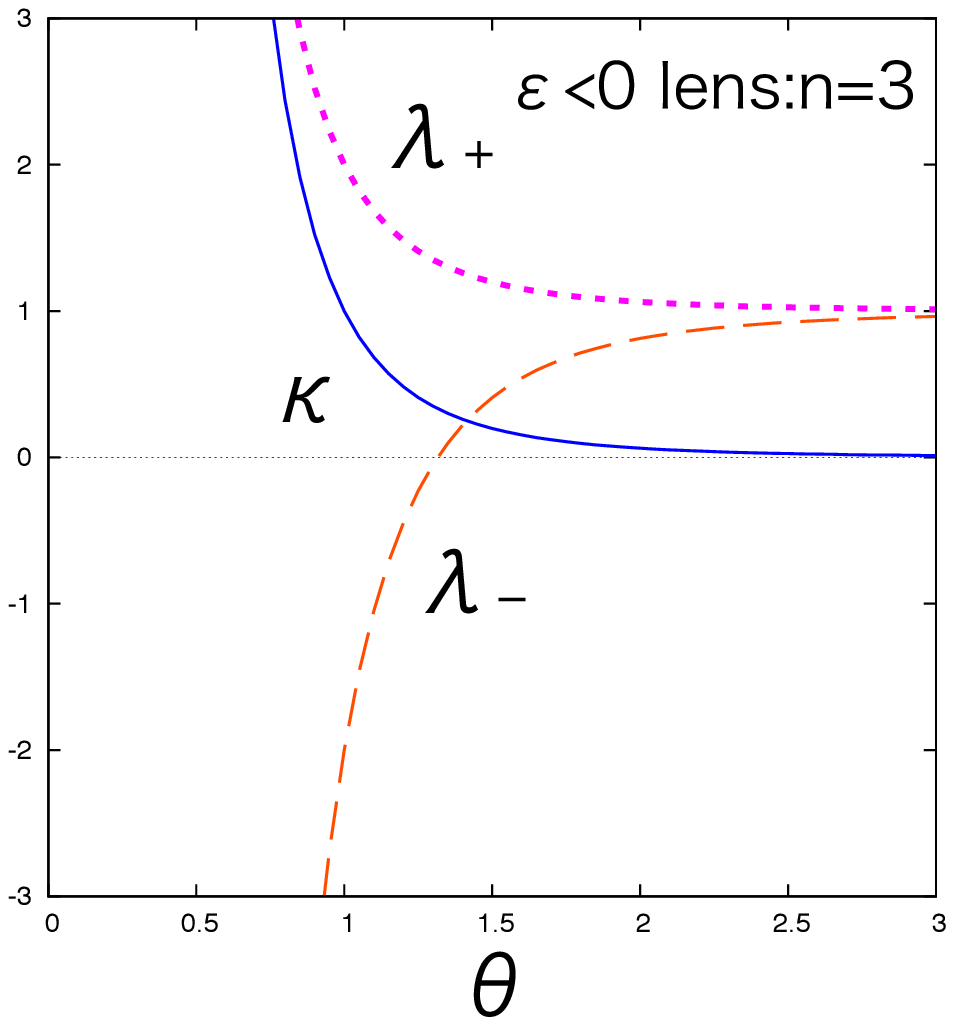}
\caption{
$\kappa$, $\lambda_{+}$ and $\lambda_{-}$ for $\varepsilon < 0$.
They are denoted by solid (blue in colors),
dotted (purple in colors) and dashed (red in colors) curves,
respectively.
The horizontal axis denotes the image position
$\theta$ in the units of the Einstein radius.
Top left: $n = 0.5$
Top right: $n = 1$.
Bottom left: $n = 2$.
Bottom right: $n = 3$.
}
\label{figure-4}
\end{figure}

By using Eq. (\ref{lenseqP2}),
we obtain
\begin{eqnarray}
\lambda_{+} = \frac{\hat\beta}{\hat\theta}
= 1 + \frac{1}{\hat\theta^{n+1}} ,
\label{lambdaP2}
\end{eqnarray}
\begin{eqnarray}
\lambda_{-} = \frac{d\hat\beta}{d\hat\theta}
= 1 - \frac{n}{\hat\theta^{n+1}} .
\label{lambdaM2}
\end{eqnarray}
One can show that $\lambda_{-} < \lambda_{+}$, if and only if $n > -1$.
Hence, both images are everywhere radially elongated.
See also Figure \ref{figure-4} for $\kappa$ and $\lambda_{\pm}$
that are numerically calculated for $n=0.5$, $1$, $2$ and $3$.
For these four cases, $\lambda_{+}$ is always larger than $\lambda_{-}$.
The convergence $\kappa$ is negative for $n=0.5$, while
it is positive for $n=2$ and $3$.
It follows that $n=1$ corresponding to the (negative-mass)
Schwarzschild lens leads to $\kappa = 0$.

Eqs. (\ref{lambdaP2}) and (\ref{lambdaM2}) give
the shear as
\begin{eqnarray}
\gamma &=& \frac{\lambda_{+} - \lambda_{-}}{2}
\nonumber\\
&=&  \frac{1+n}{2}\frac{1}{\hat\theta^{n+1}} .
\end{eqnarray}

A repulsive case might correspond to the lensing
by a void-like mass distribution.
The above calculations assume the flat (Minkowskian) background spacetime.
If one wish to consider cosmological situations,
the gravitational potential and the mass density
might correspond to the scalar perturbation and the density contrast
in the cosmological perturbation approach
based on the Friedmann-Lemaitre background spacetime \cite{SEF}.
In this cosmological counterpart, the present model with $\kappa < 0$
might correspond to an underdense region called a cosmic void,
in which the local mass density is below the cosmic mean density
and the density contrast is thus negative.
The gravitational force on the light rays by the surrounding region
could be interpreted as repulsive ($\varepsilon < 0)$,
because the bending angle of light with respect to the center
of the spherical void might be negative.
Therefore, cosmic voids might correspond to
a $\kappa < 0$ and $\varepsilon < 0$ case.
Note that the positive convergence due to the cosmic mean density
is taken into account in the definition of the cosmological distances.
There are very few galaxies in voids compared with
in a cluster of galaxies.
Hence, it is difficult to investigate gravity inside a void
by using galaxies as a tracer.
Gravitational lensing shear measurements would be another tool
for studying voids.

Before closing this section,
we mention whether we can distinguish radial elongation
and tangential one in observations without knowing the lens position.
Usually, lens objects cannot be directly seen
except for visible lens objects such as galaxies.
In particular, exotic lens models that are discussed in this paper
might be invisible.
In the above calculations, the origin of the two-dimensional
coordinates is chosen as the center of the lens object, so that
the radial and tangential directions can be well defined.
For a pair of radially elongated images ($\varepsilon < 0$),
they are in alignment with each other.
For a pair of tangentially elongated images ($\varepsilon > 0$),
they are parallel with each other.
Therefore, one can distinguish radial elongation from
tangential one by measuring such an image alignment in observations.
See also Figure \ref{figure-2} for $\varepsilon <0$ and $n=2$,
where one can see a pair of radial images.

\section{Discussion and Conclusion}
We examined gravitational lens models inspired
by modified gravity theories, exotic matter and energy.
By using an asymptotically flat, static and spherically symmetric
spacetime model of which metric depends on
the inverse distance to the power of positive $n$,
it was shown in the weak field and thin lens approximations that
images due to lens models
for the gravitational pull on light rays are tangentially elongated,
whereas those by the other models for the gravitational repulsion
on light rays are always radially distorted.

As a cosmological implication,
it is suggested that cosmic voids might correspond to
a $\kappa < 0$ and $\varepsilon < 0$ case
and hence they could produce radially elongated images
rather than tangential ones.
It would be interesting to investigate numerically
light propagation through realistic voids in cosmological
simulations, because the present model obeys a simple power-law.
It is left for future work.

We would like to thank F. Abe, M. Bartelmann, T. Harada,
S. Hayward, K. Nakao, R. Takahashi and M. Visser
for the useful conversations on the exotic lens models.
We wish to thank N. Tsukamoto for the stimulating comments
on the lecture (T.K.) at ``WormHole Workshop 2012'' in Rikkyo University.

\newpage

\begin{table}[h]
\caption{
The sign of the convergence $\kappa$.
It is the same as that of $\varepsilon (1-n)$
according to Eq. (\ref{kappa}).
}
  \begin{center}
    \begin{tabular}{c|c}
\hline
$\kappa > 0$  \; &
\;
$\varepsilon > 0 \; \& \; n <1$ \\
 \; &
\;
$\varepsilon < 0 \; \& \; n >1$ \\
\hline
$\kappa = 0$ \; &
$n =1$ \\
\hline
$\kappa < 0$ \; &
\;
$\varepsilon > 0 \; \& \; n >1$ \\
 \; &
\;
$\varepsilon < 0 \; \& \; n <1$
\\
\hline
    \end{tabular}
  \end{center}
\label{table-1}
\end{table}


\begin{thebibliography}{99}
\bibitem{shortcut}
This shortcut is not widely known.
Ref. \cite{SEF} does not mention this and does explicit calculations
of the off-diagonal terms in the Jacobian matrix
(in section 8.1 of the book).
\end{thebibliography}
\end{document}